\documentclass[aps,floatfix,showpacs]{revtex4}
\usepackage{natbib}
\usepackage{dcolumn}
\usepackage{graphicx}

\newcommand{\physrep}{Phys.~Rep.}

\newcommand{\be}{\begin{equation}}
\newcommand{\ee}{\end{equation}}
\newcommand{\bea}{\begin{eqnarray}}
\newcommand{\eea}{\end{eqnarray}}
\begin{document}
\title{A minimal set of invariants as a systematic approach to higher order gravity models}
\author{Mustapha Ishak\footnote{Electronic address: mishak@utdallas.edu}, Jacob Moldenhauer\footnote{Electronic address: jam042100@utdallas.edu}}
\affiliation{
Department of Physics, The University of Texas at Dallas, Richardson, TX 75083, USA}
\date{\today}
\begin{abstract}
Higher-order gravity models have been recently the subject of much attention in the context of cosmic acceleration. These models are derived by adding various curvature invariants to the Einstein-Hilbert action. Several studies showed that these models can have late-time self-acceleration and could, in some cases, fit various observational constraints. In view of the infinite spectrum of invariants that could be built from curvature tensors, we propose here a method based on minimal sets of independent invariants as a possible route for a systematic approach to these models. We explore a connection made between theorems on bases in invariants theory in relativity and higher-order cosmological models. A dynamical system analysis is performed and some models with accelerating attractors are discussed. The asymptotic behavior of the models is also studied using analytical calculations. 
\end{abstract}
\maketitle
\section{introduction}
Several cosmological observations \cite{observations} established that the expansion of the universe entered a phase of acceleration. The cosmic acceleration can be caused by a repulsive dark energy component that has a negative equation of state, or, a modification to gravity at cosmological scales, see for example the reviews \cite{reviews}. The cosmic acceleration and the dark energy associated with it constitute one of the most important and challenging current problems in cosmology and all physics.

As for modified gravity models, there has been recently much interest in studying higher-order gravity cosmological models. A bibliography search \cite{Biblio300} shows that nearly 300 papers studying these models have appeared in the last 3 years. The models are derived from curvature invariants that are more general than the Einstein-Hilbert action used in general relativity and standard cosmology, see for example \cite{Carroll2005,Easson2004,Easson2005} and recent reviews \cite{Lobo2008, NojiriOdintsov2006a} and references therein. Previous studies showed that the models can have some interesting dynamical features such as a late-time self-accelerating expansion without a dark energy component. It was also shown in some papers that they can have early-time inflation as well, e.g. \cite{Sotiriou2006,MengWang2004,NojiriOdintsov2003}, thereby providing a unification scenario for the two cosmic accelerating phases. In these models, the acceleration is a consequence of how the matter is coupled to the space-time. Some of the models were found to fit cosmological observations \cite{Mena2006,Shirata2005,Borowiec2006} and other models fit solar system tests \cite{NavarroVanAcoleyen2005}.  However, other papers claimed that some of the models studied fail solar system tests \cite{Chiba2003} or at least under some conditions \cite{Chiba2007}, but this has been contested in, for example, \cite{Faraoni2006a}, and also in \cite{NojiriOdintsov2007}, where models that pass these tests were discussed. In addition to the phenomenology, it has been argued that the models have theoretical motivations within unification theories of fundamental interactions and within field quantization on curved space-times \cite{BirrellDavies1982,Stelle1977}. However, previous studies stressed that the models considered were chosen somewhat randomly due to the large spectrum of possible curvature invariants \cite{Carroll2005,DologovKawasaki2003} and a systematic approach to these models is highly desirable \cite{Faraoni2006b,Sotiriou2007}. Such a systematic approach could allow one to study methodically various possibilities in this class \cite{Faraoni2006b,Sotiriou2007} and would allow one to make more general and more conclusive statements about the models.

In this paper, we propose a method based on minimal sets of invariants as a possible systematic approach to the models. We explore an idea based on theorems from the theory of invariants in general relativity \cite{Debever1964,CarminatiMcLenaghan1991,ZakharyMcIntosh1997} in order to develop a systematic framework that allows one to substantially reduce this very large class of model to a systematic list (taxonomy) of models. The idea is that curvature invariants are not independent from each other and, for a given algebraic type of the Ricci tensor (see for example the Segre classification \cite{Segre,Stephani2003}) and a given Petrov type of the Weyl tensor (i.e. symmetry classification of space-times), {\textit{e.g.}}\cite{Petrov1954,Pirani1957,Penrose1960,Stephani2003}, there exists a complete minimal independent set (basis) of these invariants in terms of which all the other invariants may be expressed. The connection between these theorems and higher-order cosmological models has not been explored before. As an immediate consequence of the proposed approach, the number of independent invariants to consider is reduced from infinity to six in the worst case and to only two in the cases of primary interest, including all Friedmann-Lemaitre-Robertson-Walker metrics. We use here the consequences of this connection in order to approach the models and study their dynamics. This should set the stage for future studies where the models will be submitted to various solar and cosmological observational constraints. 

\section{higher-order gravity models}

We refer the reader to some of the work discussing these models in some detail, see for example \cite{Carroll2005,Easson2004,Easson2005} and recent reviews  \cite{Lobo2008,NojiriOdintsov2006a} and references therein, however, let us briefly introduce the models here. In these models, various curvature scalars are usually added to the Ricci scalar into the gravitational action leading to new field equations. The new field equations change how the content of the universe is coupled to its curvature. Let us recall that the Einstein field equations and the standard Friedmann cosmology are derived from variation of the Einstein-Hilbert action:
\be
S=\frac{M_p}{2}\int{d^{4}x \sqrt{-g}R+\int{d^{4}x}\sqrt{-g}L_m}
\label{eq:HilbertEinsteinAction}
\ee
where $M_p=(8\pi G)(-1/2)$ is the reduced Planck mass, $R=g^{\alpha\beta}R_{\alpha\beta}$ is the Ricci scalar constructed from the metric tensor $g^{\alpha\beta}$ and the Ricci tensor $R_{\alpha\beta}$, and $L_m$ is the matter-energy Lagrangian. Applying variational calculus to the equation above leads to the usual Einstein equations and the usual Friedmann equations for standard cosmology:
\be
G_{\alpha\beta}=R_{\alpha\beta}-\frac{1}{2}R\,g_{\alpha\beta}=\frac{1}{M_p^2}T_{\alpha\beta}.
\label{eq:EinsteinEquations}
\ee
The higher-order invariant models under discussion have a more general action of the form
\be
S=\frac{M_p}{2} \int{d^{4}x \sqrt{-g} f(R,R^{\alpha\beta} R_{\alpha\beta},R^{\alpha\beta\gamma\delta} R_{\alpha\beta\gamma\delta},R^{\alpha\gamma} R_{\alpha\beta} R^{\beta}_{\gamma},R^{\alpha\beta\mu\nu} R_{\alpha\beta\gamma\delta} R^{\gamma\delta}\,_{\mu\nu},...)}+\int{d^{4}x \sqrt{-g}L_m}
\label{eq:HilbertEinsteinAction}
\ee
where $R_{\alpha\beta\gamma\delta}$ is the Riemann curvature tensor and $f$ is a function of the curvature invariants. It can be quickly realized that there are an infinite number of curvature invariants that one can construct from combinations of the metric tensor, the Ricci tensor, and the Riemann tensor and their powers. 

Relevant to the question of cosmic acceleration, it has been shown in previous studies  that some models with inverse powers of these scalars have negligible contribution to the dynamics at the proximity of strong gravitational fields but then start contributing significantly to the dynamics at very large (cosmological) scales producing an accelerating expansion. For example many papers have focused on inverse powers of $R$ and also in some cases inverse powers of combinations of $R$, $R^{\alpha\beta} R_{\alpha\beta}$ and $R^{\alpha\beta\gamma\delta} R_{\alpha\beta\gamma\delta}$, e.g. \cite{Carroll2005,Mena2006}. 

\section{Minimal sets of curvature invariants and higher order gravity models}
Curvature invariants are important in general relativity studies since they allow a manifestly invariants characterization of some properties of spacetimes. There has been some pioneering work by \cite{Thomas1934,NarlikarKarmarkar1948,GeheniauDebever1956,Debever1956,Debever1964,Witten1959,Petrov1969} and then a renewed interest by several  recent papers \cite{CarminatiMcLenaghan1991,ZakharyMcIntosh1997,Santosuosso1999,CarminatiZakhary1999}. This long series of studies resulted in theorems about minimal sets of invariants and classifications of spacetime manifolds. It was shown for example in \cite{CarminatiMcLenaghan1991,ZakharyMcIntosh1997} that there are {\textit {at most}} 14 independent real algebraic curvature invariants in a 4-dimensional Lorentzian space. The number of independent invariants depends on the symmetries of the spacetime as delineated by the Petrov classification \cite{Petrov1954,Pirani1957,Penrose1960,Stephani2003} and also the algebraic type of the Ricci tensor as for example described by the Segre classification \cite{Segre,Stephani2003}). The other invariants of the spacetime can be written in terms of this complete minimal set (basis).   

The Petrov classification is based on the algebraic structure of the Weyl curvature tensor $C_{\alpha\beta\gamma\delta}$. In an n-dimensional spacetime, the Weyl tensor is related to the Riemann and Ricci tensors by 
\be
C_{\alpha\beta\gamma\delta}=R_{\alpha\beta\gamma\delta}+(g_{\alpha\delta}R_{\gamma\beta}+g_{\beta\gamma}R_{\alpha\delta}-g_{\alpha\gamma}R_{\beta\delta}-g_{\beta\delta}R_{\alpha\gamma})/(n-2)+R\,(g_{\alpha\gamma}g_{\beta\delta}-g_{\alpha\delta}g_{\beta\gamma})/(n-1)(n-2)
\label{eq:Weyl}
\ee
The Petrov symmetry classification types are noted by \{Type I, II, D, III, N, O\} and can be understood in the following way. One can consider the Weyl tensor as a fourth-rank tensor operator acting on bi-vectors of space-time. Like in linear algebra, the eigenvalues and eigenbivectors will have some multiplicities. These multiplicities indicate symmetries of the Weyl tensor acting as an operator and also the symmetries of the space-time. These multiplicities are also related to the principal null directions and determine the Petrov type of the space-time. The Petrov classification's theorems give, for example, type I as the most general case with four independent null directions and type O as the simplest case. Current studies, using additions of invariants to the Einstein-Hilbert's action, have been done using the flat Friedmann space-times. With the approach that we propose, the simplest case (Type O) still covers all the  Friedmann-Lemaitre-Robertson-Walker metrics.

Next, the Segre classification is based on the algebraic structure of the trace-free Ricci tensor given by 
\be
S_{\alpha\beta}=R_{\alpha\beta}-\frac{R}{4}g_{\alpha\beta},
\label{eq:Trace-Free-Ricci}
\ee 
where the tensor is viewed as a second-rank operator acting on space-time vectors. The eigenvalue equation 
\be
S^{\alpha}\,_{\beta}v^{\beta}=\lambda v^{\alpha}
\ee 
is then considered and the classification is based on the multiplicity of the eigenvalues and also whether the eigenvectors are null, timelike or spacelike (see e.g. Chap. 5 in \cite{Stephani2003}). For example, the Friedmann-Lemaitre-Robertson-Walker metrics are of Segre type A1-$[(1\,1\,1,\,1)]$ with two eigenvalues, one with multiplicity 1 and the other with multiplicity 3 (three equal trace-free Ricci components) and this is the type of interest that we consider in the next section.  

It is worth clarifying that Petrov and Segre classifications depend on the symmetries of the space-time metric under consideration and were originally derived without reference to source fluids. Indeed, most papers as for example \cite{GeheniauDebever1956,Debever1956,Debever1964,ZakharyMcIntosh1997,Santosuosso1999} derived and discussed minimum sets of invariants in terms of the algebraic Segre and Petrov types without reference to source fluids, while \cite{CarminatiMcLenaghan1991} considered the type of source fluid in their discussion as obtained via the Einstein field equations. Since we study here modified gravity models where the field equations are not anymore the Einstein equations, we thus consider the minimum sets from the point of view of the algebraic Segre and Petrov classifications of the space-time manifold. 

Now, as it was derived in original papers on curvature invariants, the elements of the basis can be built from contractions of the trace-free Ricci tensor
plus those of the Weyl tensor above, $C_{\alpha\beta\gamma\delta}$, as given by equation (\ref{eq:Weyl}) and the complex conjugate of its  self-dual tensor, i.e. $\bar{C}_{\alpha\beta\gamma\delta}$ \cite{CarminatiMcLenaghan1991,ZakharyMcIntosh1997}.

The first element of the basis is the Ricci scalar, $R$. The following elements are usually put into three categories. The elements that are built from contraction of trace-free Ricci tensors only are noted with names starting with an $r$ in the notation of \cite{CarminatiMcLenaghan1991} or with an $E$ in the notation of \cite{GeheniauDebever1956}, as for example
\be
r_1=\frac{1}{4}E_{(1)}=R1= \frac{1}{4}S_{\alpha}\,^{\beta}\, S_{\beta}\,^{\alpha}. 
\label{eq:R1}
\ee
Then there are pure Weyl elements starting with $w$ in the notation of \cite{CarminatiMcLenaghan1991} or with an $C$ in the notation of \cite{GeheniauDebever1956}, such as 
\be
w_1= W1=\frac{1}{4}\bar{C}_{\alpha\beta\gamma\delta}\,\bar{C}^{\alpha\beta\gamma\delta},
\label{eq:W1}
\ee
and mixed elements starting with $m$ in the notation of \cite{CarminatiMcLenaghan1991} or with an $D$ in the notation of \cite{GeheniauDebever1956}, as for example,
\be
m_1=M1=\frac{1}{4}\bar{C}_{\alpha\gamma\beta\delta}\,S^{\gamma\delta}\,S^{\alpha\beta}.
\label{eq;M1}
\ee
We use here a notation similar to, for example, Table II in \cite{CarminatiMcLenaghan1991} except that we use $R1$ uppercase with no-subscript instead of $r_1$ as they did. We chose this small change of notation in order to be close to the recent notation using capital letter for invariants used to construct higher order gravity models e.g. \cite{Carroll2005} such as $P=R^{\alpha\beta} R_{\alpha\beta}$ and $Q=R^{\alpha\beta\gamma\delta} R_{\alpha\beta\gamma\delta}$. 

These previous studies on invariant theory were focused on building minimal sets in mathematical relativity with no reference to  higher-order gravity models. We study here the implications of this connection with higher-order cosmological models and their dynamics.   

\section{Higher-order gravity models based on minimal sets}

Aiming to proceed in a systematic way, we consider a Petrov classification of type O (the simplest) and a Segre type A1-$[(111),1]$ \cite{Segre,Stephani2003}. As mentioned earlier, this case includes all the Friedmann-Lemaitre-Robertson-Walker manifolds. In this case the basis of invariants reduce to only two elements, see for example \cite{ZakharyMcIntosh1997,CarminatiMcLenaghan1991,GeheniauDebever1956}, namely 
\be
\{R,\,R1\}
\ee
or similarly $\{R,E_{(1)}\}$. A subtle point here is that the invariant $R1=\frac{1}{4}S_{\alpha}\,^{\beta}\, S_{\beta}\,^{\alpha}$ is built from the trace-free Ricci tensor and is a true second element for the basis since it is independent from $R$, unlike the commonly used $P=R^{\alpha\beta} R_{\alpha\beta}$. Indeed, the trace part, $R$, is removed from $R1$ but this is not the case for $P$. The consequences of this are reflected as a significant difference between the dynamical equations that are derived respectively from models built using $P$ or using $R1$, in the sense that $\{R,R1\}$ provides a simpler and independent set of building blocks to construct such theories. In order to illustrate this, let's consider the flat Friedmann-Lemaitre-Robertson-Walker metric, 
\be
ds^2=-dt^2+a(t)^2 d\vec{x}^2,
\label{eq:FLRWflatmetric}
\ee
where $a(t)$ is the scale factor. Next, we calculate the contribution of $-\frac{m^6}{R1}$ to the generalized Friedmann equation and find 
\be
-2\frac{m^6}{\dot{H}^4}(\dot{H}^2-2H^2\dot{H}+2H\,\ddot{H})
\label{eq:FriedmannR1}
\ee
where $m$ is defined in the usual way to have dimensions of mass, an overdot signifies taking the derivative with respect to cosmic time, and $H(t)=\frac{\dot{a(t)}}{a(t)}$ is the Hubble parameter used to write the third order differential equation in a(t) as a second order differential equation in $H(t)$. Now, for the $-\frac{m^6}{P}$ that was previously used in the literature, the addition to the Friedmann equation is given by 
\be
-\frac{m^6}{8(3H^4+3H^2\dot{H}+\dot{H}^2)^3}(\dot{H}^4+11H^2\dot{H}^3+2H\dot{H}^2\ddot{H}+33H^4\dot{H}^2+30H^6\dot{H}+6H^3\dot{H}\ddot{H}+6H^8+4H^5\ddot{H}).
\label{eq:FriedmannP}
\ee
One can see immediately that the dynamical equation from $R1$ is significantly reduced compared to the one built from $P$. Equations (\ref{eq:FriedmannR1}) for $R1$ is free from any redundancy present in equation (\ref{eq:FriedmannP}) for $P$. While the two constructions lead to different theories, the basis $\{R,R1\}$ provides the  simplest independent set of building blocks to construct systematically such theories. Similar simplifications are expected for other higher-order invariants models in the taxonomy. 

Next, we derive the general field equations for $f(R,R1)$ models using the action of the form
\be
I=\frac{M_p}{2}\int{d^{4}x \sqrt{-g}[R+f(R,R1)+\int{d^{4}x}\sqrt{-g}L_m}
\label{eq:ActionRR1}
\ee
that we vary with respect to the metric and we find
\bea
\lefteqn{S^{\alpha\beta}-\frac{1}{4}g^{\alpha\beta}R-\frac{1}{2}g^{\alpha\beta}f+f_{R}S^{\alpha\beta}+\frac{1}{4}f_{R}g^{\alpha\beta}R+g^{\alpha\beta}f_{R;\gamma}\,^{\gamma}-f_{R;}\,^{\alpha\beta}+\frac{1}{2}f_{R1}S^{\alpha\gamma}S^{\beta}\,_{\gamma}+\frac{1}{8}f_{R1}S^{\alpha\beta}R}\nonumber\\
& &+\frac{1}{4}(f_{R1}S^{\alpha\beta})_{;\gamma}\,^{\gamma}+\frac{1}{4}g^{\alpha\beta}(f_{R1}S^{\gamma\delta})_{;\gamma\delta}-\frac{1}{4}(f_{R1}S^{\gamma\beta})_{;}\,^{\alpha}\,_{\gamma}-\frac{1}{4}(f_{R1}S^{\gamma\alpha})_{;}\,^{\beta}\,_{\gamma}=8\pi GT^{\alpha\beta}
\label{eq:FieldEquationsRR1}
\eea
where we have used the definitions $f_{R}\equiv\frac{\partial{f}}{\partial{R}}$ and $f_{R1}\equiv\frac{\partial{f}}{\partial{R1}}$. We study the dynamics of some $f(R,R1)$ models in the next sections. 

\section{model 1: $f(R,R1)=-\frac{m^{4n+2}}{\alpha(\frac{1}{6}R^2-8R1)^n}$}

Let's construct some examples based on the minimal set $\{R,R1\}$. We use some guidance from previous studies \cite{Stelle1977,Chiba2005,Dvali2006,LiBarrowMota2007,NavarroVanAcoleyen2006,DeFelice2006} in order to satisfy some conditions for building models with physical degree of freedom. As discussed in these papers, a theory with action $R+f(GB)$ can be re-written as the Einstein-Hilbert action plus a $GB$-function coupled to a scalar field $\phi$ with potential $U(\phi)$, i.e. $R + f(\phi) GB-U(\phi)$ \cite{DeFelice2006}. In the latter frame, the theory becomes like that of a Gauss-Bonnet one where the equations of motion (EOMs) then decouple into second order equations for the metric and for each scalar field involved. For example, \cite{DeFelice2006} derived some conditions for GB-like invariants so that the EOMs do decouple into second order equations for the metric and each scalar field. It is worth clarifying that the decoupled equations are not the background field equations in the $R+f(GB)$ frame where the Friedmann equation (the 00-field equation) contains third derivatives of the metric. As discussed in \cite{DeFelice2006}, the cancellation of fourth order derivatives in the decoupled equations is a necessary condition to have a theory free from massive spin-2 gravitons (ghosts)\cite{Stelle1977,Chiba2005,Dvali2006,NavarroVanAcoleyen2006,DeFelice2006}. The condition imposes for us some relations between the parameters that we can use in order to combine $R$ and $R1$. The first example we consider is where the function is that of the pure GB invariant given by, e.g. \cite{Stelle1977,Chiba2005,NavarroVanAcoleyen2006,DeFelice2006},
\be
GB=R^2-4R^{\alpha\beta}R_{\alpha\beta}+R^{\alpha\beta\gamma\delta}R_{\alpha\beta\gamma\delta}=R^2-4P+Q
\label{eq:GaussBonnet} 
\ee                                                 where we can recall the earlier conventional definitions for these scalars $P=R^{\alpha\beta} R_{\alpha\beta}$ and $Q=R^{\alpha\beta\gamma\delta} R_{\alpha\beta\gamma\delta}$.             
Now using the identity 
\be
C^{\alpha\beta\gamma\delta}C_{\alpha\beta\gamma\delta}=C^2= \frac{1}{3}R^2-2P+Q,
\label{eq:Weylsquared}
\ee
setting $C^2=0$ for Friedmann manifolds, and writing the result in terms of $\{R,R1\}$ reduces the $GB$ invariant to $\frac{R^2}{6}-8R1$. We are interested in inverse powers of these invariants and, in a systematic way, the first example to explore should be
\be
f(R,R1)=-\frac{m^6}{\frac{R^2}{6}-8R1}.
\ee
Now, to investigate if the model has a late time self-acceleration phase, we perform a dynamical system analysis. Following previous studies, we simplify the study to vacuum solutions to study asymptotic behavior of the universe since the inclusion of matter at late times as dust will not significantly contribute to the dynamics. 

However, one may question if the separatrix analysis done further below between matter dominance and vacuum dominance will not be affected by such a simplification. This question was addressed in \cite{Carroll2005} where it was shown that adding matter to the vacuum analysis will not affect the fixed points for power-law solutions. It is also reassuring that some of our results on the separatrix analysis using dynamical systems for vacuum are consistent with results from other studies \cite{DeFelice2006} that used analytical derivations and showed that such separatrix are present in similar models. Nevertheless, it remains of interest to verify this assumption using our approach in future work.

Now, for the Friedmann equation from variation of $R+f(R,R1)$) reads
\be
3H^2-\frac{m^6}{24H^2(\dot{H}+H^2)^3}(H^4+6H^2\dot{H}+H\ddot{H}+3\dot{H}^2)=0
\label{eq:FriedmannGB1}
\ee 

\begin{figure}
\begin{center}
\begin{tabular}{|c|c|}
\hline

{\includegraphics[width=2.8in,height=2.8in,angle=0]{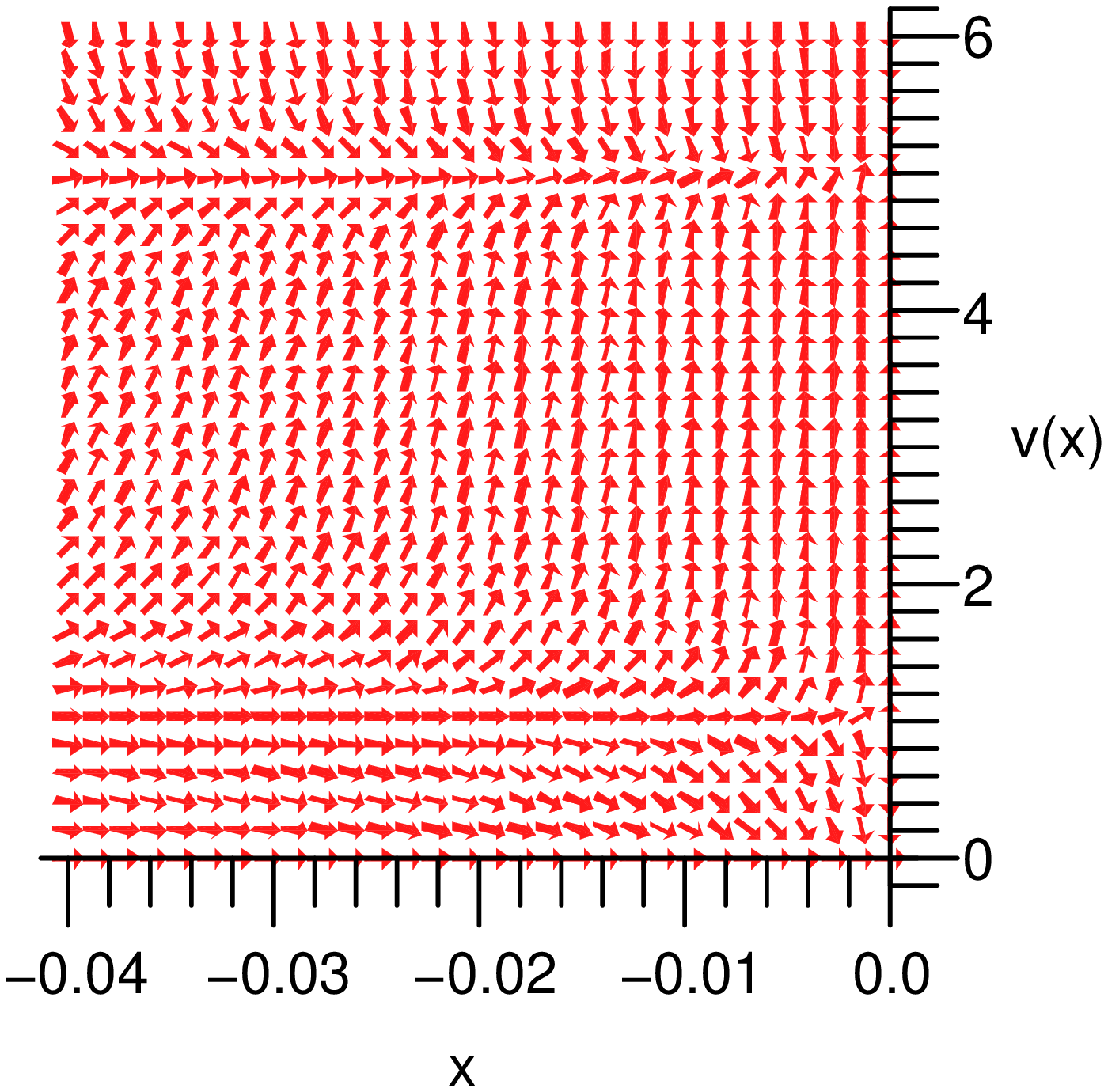}} &
{\includegraphics[width=2.8in,height=2.8in,angle=0]{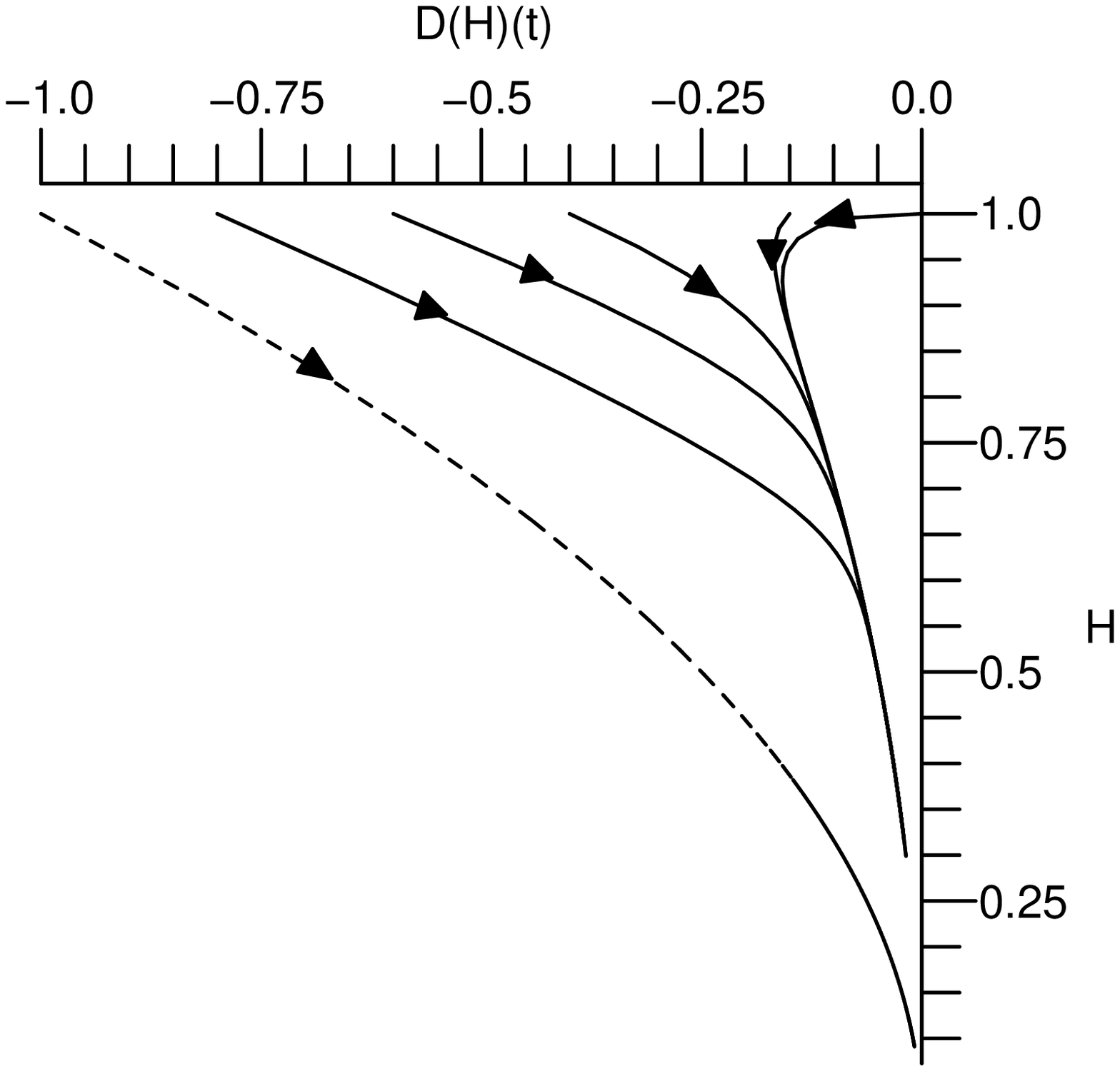}} \\
\hline
\end{tabular}
\caption{
LEFT: Dynamical system analysis for the case study $f(R,R1)=-\frac{m^6}{(R^2/6-8R1)}$. In the first order phase portrait of the coordinate plane $(x,v)$, we find an accelerating attractor with the scale factor a(t) evolving as  $t^p$ where $p= 5$.  There is also a repeller at $p=1$.
RIGHT: In the second order phase portrait with coordinates $(\dot{H},H)$, the solid-line branches are the accelerating ones with attractor having the scale factor a(t) as indicated on the left. The dashed line is the separatrix solution where $-\dot{H}/H^2=1$ and corresponds to the repeller at $p=1$. The horizontal axis on the right is $dH/dt$. 
} 
\end{center}
\end{figure}

As usual, see for example \cite{Carroll2005}, we parameterize the second order differential equation in $H(t)$ to a first order differential equation in $y(x)$ by setting $x=-H(t)$, $y=\dot{H}(t)$, so $\ddot{H} =-y\, dy/dx$ by expanding the differential by separation of variables in $d\dot{H}/dt$ and perform a phase portrait analysis of the Friedmann equation. For cosmologically significant accelerating solutions, we examine power-law solutions that take the form $a(t)\propto t^p$ in which we examine the relationship $\dot{H}=-H^2/p$ or $y=x^2/p$.  In the same way, we define the asymptotic function $v(x)$ as 
\be
v(x)=-\frac{x^2}{y}, \,or\,y=-\frac{x^2}{v},
\label{eq:v(x)parameter}
\ee
with $v\not=0$ so that
\be
dy=\frac{-2x\,v\,dx+x^2dv}{v^2}.
\label{eq:diffv(x)parameter}
\ee
We draw attention to a strong point in \cite{Carroll2005}, if $\left|\dot{H}\right|=\left|y\right|\rightarrow{\infty}$, then $\left|v\right|\rightarrow{0}$ if $x\not=0$.  Clearly, this means with $x$ non-zero, as $v\rightarrow{0}$ we anticipate the singularity $\left|\dot{H}\right|\rightarrow{\infty}$. We can now convert equation (\ref{eq:FriedmannGB1}) into a first order Friedmann equation 
\be
x\frac{dv}{dx}=5v-6v^2+v^3+\frac{x^6(72-216v+216v^2-72v^3)}{m^6}
\label{eq:Friedmannv(x)GB1a}
\ee
which allows us to study the asymptotic behavior of the solutions to this equation at late times as seen in the phase portrait on the left side of Fig. 1. The attracting solutions for the asymptotic future are close to $x=0$ and at earlier times in the evolution of the universe lie in the negative values of $x$ because the earlier parameterization of $x=-H(t)$ denotes the x-axis as the negative of the Hubble parameter. However, since we are interested in asymptotically vacuum solutions, we can analytically solve for these late time solutions, as well. For this, we solve for $v(x)$ as $x\rightarrow{0}$ in (\ref{eq:Friedmannv(x)GB1a})
\be
0=v_{0}(5-6v_{0}+v_{0}^2)
\label{eq:GB1v0}
\ee
noticing that as $H\rightarrow{0}$, $v_{0}\rightarrow{p}=\,$constant in equations (\ref{eq:v(x)parameter}) allows us to see that the asymptotic solutions of equation (\ref{eq:GB1v0}) have constant values of $v_{0}=5, 1, 0$ corresponding to power-law expansion with an attractor $p=5$ and a repeller at $p=1$ as seen in Fig. 1.  The solution for $p=v_0=0$ corresponds to $y=\dot{H} \rightarrow{\infty}$ which is a singularity, also discussed in \cite{Carroll2005}. The attractor needs to be greater than $1$ for accelerating expansion (this can be seen directly from using $a(t)\propto t^p$ into the deceleration parameter (\ref{eq:decelerationparameter})), so we assume some initial conditions ($\ddot{a}>0$) will eliminate the universe finding itself at solutions of $p=0, 1$.  In fact, these last two solutions are singular points for the generalized Friedmann equation (\ref{eq:FriedmannGB1}) and the presence of a singularity at $p=1$ is problematic, because as we discussed earlier, due to the equivalence of matter and vacuum fixed point solutions as shown in \cite{Carroll2005}, which we have assumed throughout this paper, this solution of $p$ represents the transition point from matter domination (deceleration) to the accelerating phase, i.e. the point where $\ddot{a}=0$. If this point  cannot be crossed, it means that there is a separatrix at this solution, rendering the action cosmologically not viable.  Both singularities are evident from the Friedmann equation (\ref{eq:FriedmannGB1}) second order in $H(t)$, where, the denominator equals zero for 
\be
\dot{H}+H^2=0
\label{eq:GB1Zero}
\ee
where using power law solutions, $a(t)=t^p$, translates into
\be
\frac{p(-1+p)}{t^2}=0
\label{eq:GB1ZeroP}
\ee
and confirms the singular points $p=0,1$. The singularities, if any, will come from any real solutions to equations equivalent to (\ref{eq:GB1ZeroP}). The singular point $p=1$, as confirmed by our earlier dynamical system approach was identified as a repeller. This is also a separatrix not allowing for the transition from matter domination to an accelerating phase. Where as, the attractor at $p=5$ shows that the model with inverse powers of the pure GB action expressed in the basis $\{R,R1\}$ do have the desired accelerating late-time behavior, we should try to find actions without this separatrix and transition problem. 

In looking at the more traditional second-order phase portrait in the coordinate plane of $(\dot{H},H)$ we can also see the attractor in Fig. 1.  The solid lines leading to the accelerating attractor and the dotted line for the non-accelerating solution.   More explicitly, in this Einstein-frame we can consider the analysis of the deceleration parameter \cite{Rindler2006} $q$
\be
q=-\frac{1}{H^2}(\dot{H}+H^2)=-1-\frac{\dot{H}}{H^2}
\label{eq:decelerationparameter}
\ee
where due to gravitational attraction it was thought that the universe was slowing down in its expansion at the time $q$ was defined, a positive $q$ meant deceleration.  So, for $q<{0}$ we have acceleration as $-1-\dot{H}/H^2<{0}$, or we can define a condition for the accelerating expansion as
\be
-\frac{\dot{H}}{H^2}<{1}
\label{eq:accelerationparameter}
\ee
Using this \textit{acceleration} condition in the phase portrait of the right side of Fig. 1, we can see that coordinates of $(\dot{H},H^2)<{1}$ will be leading to accelerating attractors and coordinates of $(\dot{H},H^2)>{1}$ will be non-accelerating. We can now see in Fig. 1 that the dashed line corresponds to the separatrix, $-\dot{H}/H^2=1$ because the solutions above this curve would be accelerating corresponding to the $p=5$ attractor, and any solutions below this curve would be non-accelerating of which we have none for this model.  Also, notice that the de Sitter solution is at $(0,1)$.   

In looking to eliminate other models that contain the $\ddot{a}=0$ separatrix, we continue in a systematic way, by generalizing this same analysis as actions with pure GB forms written in the basis $\{R,R1\}$ to  
\be
f(R,R1)=-\frac{m^{4n+2}}{\alpha(\frac{1}{6}R^2-8R1)^n},
\label{eq:Generalf(R,R1)Action2}
\ee
where $\alpha$ is a dimensionless constant and $n>0$ is an integer, for both phase spaces, for which the general Friedmann equations for models of this type become
\bea
\lefteqn{3H^2-\frac{m^{4n+2}}{2\alpha (24H^2)^n(\dot{H}+H^2)^{2+n}}[6nH^2\dot{H}+4n^2H^2\dot{H}}\nonumber\\
& &+\dot{H}^2+2H^2\dot{H}+H^4+nH\ddot{H}+nH^4+n^2H\ddot{H}+2n^2\dot{H}^2+3n\dot{H}^2]=0.
\label{eq:GeneralGB1Freidmann}
\eea
We can now convert equation (\ref{eq:GeneralGB1Freidmann}) into a first order generalized Friedmann equation 
\be
x\frac{dv}{dx}=\frac{-v}{n(n+1)}[6nv+4n^2v-1+2v-v^2-5n-nv^2-4n^2]-\frac{v(6\alpha-12v\alpha+6v^2\alpha)x^2}{m^{4n+2}n24^{-n}(\frac{x^4(-1+v)}{v})^{-n}(n+1)}
\label{eq:GeneralFriedmannV(x)1}
\ee
We can analytically solve for the late-time solutions, in this general case as well. For this, we solve for $v(x)$ as $x\rightarrow{0}$ in (\ref{eq:GeneralFriedmannV(x)1})

\be
0=-\frac{v_{0}(6nv_{0}+4n^2v_{0}-1+2v_{0}-v_{0}^2-5n-nv_{0}^2-4n^2)}{n(n+1)}
\label{eq:V(x)GeneralSolution}
\ee
This asymptotic approach yields constant values of $v_{0}=0,1,4n+1$, where we have seen a specific study of the case where $n=1$, (see Fig.1) allows us to determine that these solutions correspond to an accelerating attractor at $p=4n+1$, a singularity at $p=0$ and a singular repeller at $p=1$. For confirmation of the separatrix, we study the denominator of the Friedmann equation (\ref{eq:GeneralGB1Freidmann}), to find the singular points where the denominator vanishes.  Again, even in the generalized pure GB model, the nontrivial part for this study comes from $\dot{H}+H^2=0$ which is solved in the same way for the singular point $p=1$, as confirmed by our earlier dynamical system approach. 

So, we see that these pure GB models have a late-time accelerating power-law attractor as desired however they also have a separatrix and thus fail to provide a passage from matter domination to late-time self acceleration. We have shown this separatrix explicitly here using dynamical systems in our basis $\{R,R1\}$, a result that is consistent with previous claims \cite{DeFelice2006}. In order to avoid this limitation, we will look in the next section at non-pure GB models.  
\begin{figure}
\begin{center}
\begin{tabular}{|c|c|}
\hline

{\includegraphics[width=3.0in,height=3.0in,angle=0]{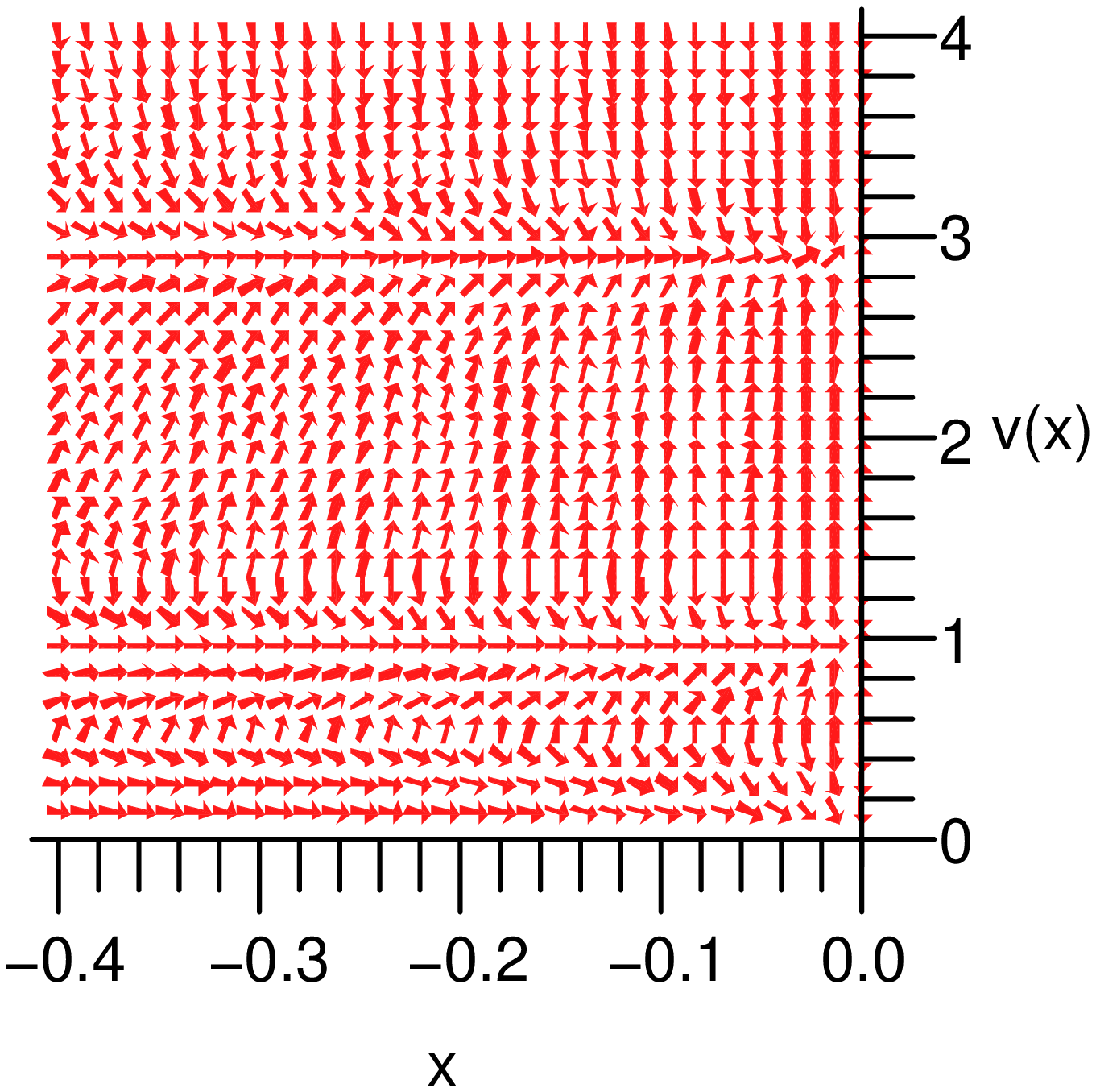}} &
{\includegraphics[width=3.0in,height=3.0in,angle=0]{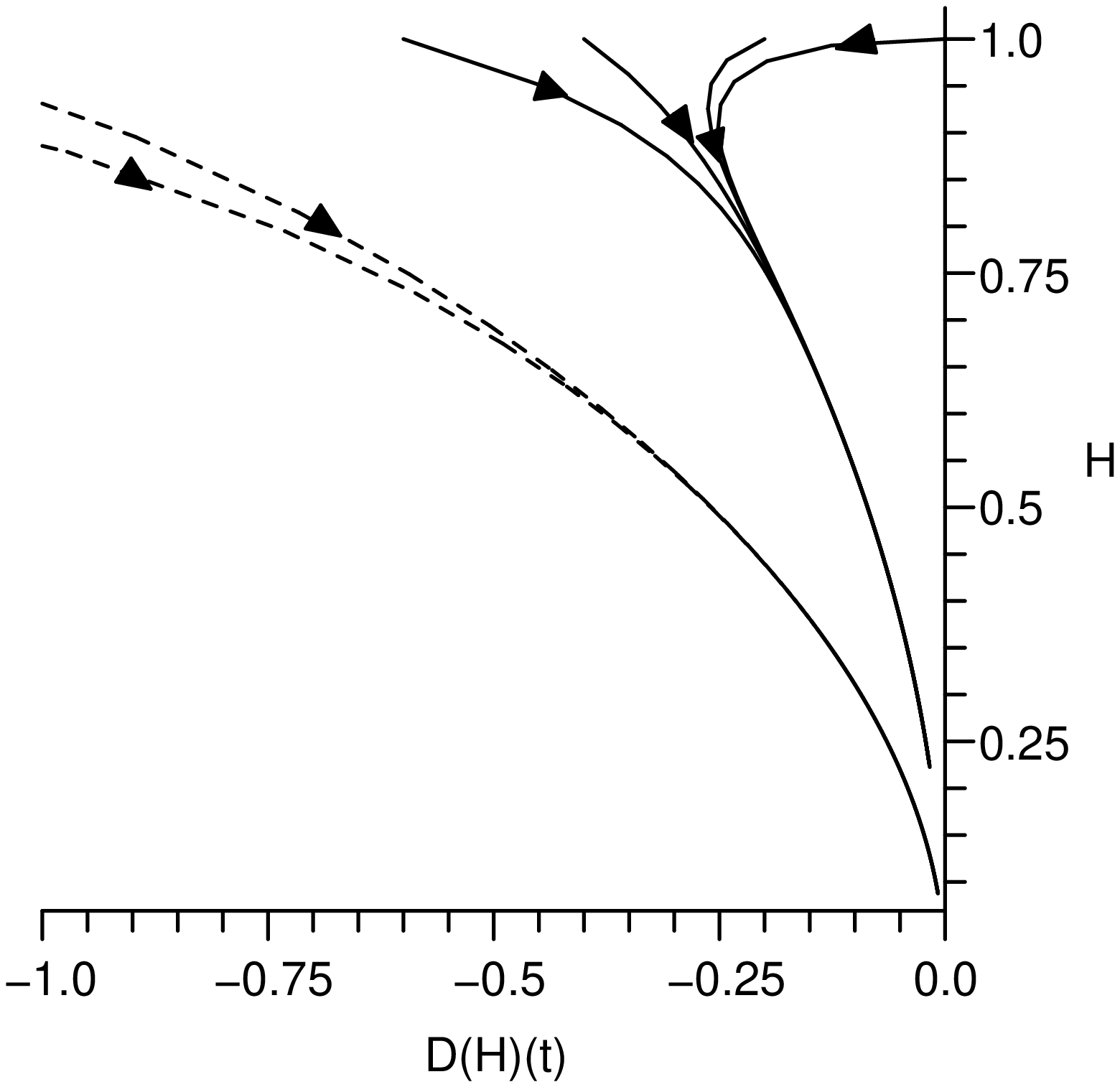}} \\
\hline
\end{tabular}
\caption{ 
LEFT: Dynamical systems for the case study $f(R,R1)=\frac{m^6}{R^2/3+8R1}$. In the first order phase portrait of the coordinate plane $(x,v)$, we find an accelerating attractor with the scale factor a(t) evolving as  $t^p$ where $p\approx2.9078$ and the non-accelerating attractor with $p\approx0.96725$.  
RIGHT: In the second order phase portrait with coordinates $(\dot{H},H)$, the solid-line branches are the accelerating ones with attractor having the scale factor a(t) as indicated on the left. The dashed lines are non-accelerating. The horizontal axis on the right is $dH/dt$. 
} 
\end{center}
\end{figure}
\section{Model 2: $f(R,R1)=\frac{m^6}{\frac{1}{3}R^2+8R1}$}

Using the basis $\{R,R1\}$, we re-express here models based on the general invariant  obtained from
\be
a_1R^2+a_2P+a_3Q
\label{eq:GeneralGBForm}
\ee
where, as discussed in \cite{DeFelice2006} (see the begining of our previous section), the fourth-order derivatives in the decoupled equations are eliminated by using the combination $a_2=-4a_3$ \cite{DeFelice2006} of the dimensionless constants, $a_1$, $a_2$, and $a_3$. In the previous section, we considered models with functions of the pure GB invariant where we also required $a_1=a_3$ but here we relax this condition. We use again (\ref{eq:Weylsquared}) with $C^2=0$ to derive in terms of the basis $\{R,\,R1\}$ the combination
\be
a_1R^2+a_3(-\frac{5}{6}R^2-8R1)
\label{eq:GeneralGB1Form}
\ee
where we can see by setting $a_1=a_3$ we return to the pure GB form.  So, we must set $a_1 \ne a_3$ to avoid the separatrix at $\ddot{a}=0$ described in previous section.  For example, we choose $a_3=1$ and $a_1=1/2$ to build the inverse action 
\be
f(R,R1)=\frac{m^6}{\frac{1}{3}R^2+8R1}
\label{eq:GB1ModelA1=1/2}
\ee
where the minus sign in front of the function was absorbed into the denominator. Variation of this action with respect to the metric will lead to equations of motion that give the following Friedmann equation 
\bea
\lefteqn{3H^2+\frac{m^6}{12(3\dot{H}^2+8H^2\dot{H}+8H^4)^3}(27\dot{H}^4+258H^2\dot{H}^3+744H^4\dot{H}^2+624H^6\dot{H}}\nonumber\\
& &+128H^8+54H\dot{H}^2\ddot{H}+144H^3\dot{H}\ddot{H}+80H^5\ddot{H})=0
\label{eq:FriedmannModelA1=1/2}
\eea

The phase portrait for equation (\ref{eq:FriedmannModelA1=1/2}) in the coordinate space $(\dot{H},\,H)$ is shown in Fig. 2, where we can see two attractors.  With the \textit{acceleration} condition we can see in the phase portrait of the right side of Fig. 2, that coordinates of $(\dot{H},H^2)<{1}$ will be leading to accelerating attractors and coordinates of $(\dot{H},H^2)>{1}$ will be non-accelerating attractors.  The solid lines lead to the accelerating attractor, and the dashed lines lead to the non-accelerating attractor. Also, notice that the de Sitter solution is at $(0,1)$. 

For the study of power-law solutions in $a(t)\propto{t^p}$ we convert equation (\ref{eq:FriedmannModelA1=1/2}) using the earlier parameterization in $y(x)$.  The Friedmann equation now reads
\bea
x \frac{dv}{dx} & = & \frac{135v^2-546v^3+940v^4-624v^5+128v^6}{2v(27-72v+40v^2)} \nonumber\\
& &+\frac{(972-7776v+28512v^2-59904v^3+76032v^4-55296v^5+18432v^6)x^6}{2m^6v(27-72v+40v^2)}
\label{eq:FriedmannV(x)ModelA1=1/2}
\eea
Its solutions in the space of $(x,\,v)$ can be seen in Fig. 2. As an analytical confirmation of these results, we study asymptotic solutions as $x\rightarrow{0}$ in equation (\ref{eq:FriedmannV(x)ModelA1=1/2}),
\be
0=\frac{135-546v_{0}+940v_{0}^2-624v_{0}^3+128v_{0}^4}{2(27-72v_{0}+40v_{0}^2)}
\label{eq:FriedmannV(0)ModelA1=1/2}
\ee
where we eliminated the trivial solution. Solving this equation for $v_{0}=p=constant$ gives two real solutions and two complex solutions.  Only the real solutions, $v_{0}=31/16+\sqrt{241}/16$ and $v_{0}=31/16-\sqrt{241}/16$, are of interest for power-law attractors corresponding to $p\approx2.9078$ and $p\approx0.96725$,  respectively. Again, only the attractor $p>1$ can be of interest for late-time acceleration, so we would need to assume some initial conditions to avoid the non-accelerating solution.  But the other attractor could serve to be useful to fit the universe to some non-accelerating phases allowing for structure formation, see e.g.  \cite{Carroll2005}. A check to make sure that we have avoided all the singular points for this model needs to be performed.  A separatrix in the evolution of the universe between radiation domination, $p=1/2$, and the transition point, $p=1$ would keep the universe from reaching its accelerating phase, so we must make sure there are no singular points within this interval for this model.  As before, we need to determine solutions to the denominator of the generalized Friedmann equation (\ref{eq:FriedmannModelA1=1/2})
\be
3\dot{H}^2+8H^2\dot{H}+8H^4=0.
\label{eq:FriedmannModelA1=1/2Zero}
\ee
Now, we transform (\ref{eq:FriedmannModelA1=1/2Zero}) in the same way as the previous section to write
\be
\frac{p^2(3-8p+8p^2)}{t^4}=0
\label{eq:FriedmanModelA1=1/2ZeroP}
\ee
Solving the nontrivial part of equation (\ref{eq:FriedmanModelA1=1/2ZeroP}) yields two complex solutions. Since there are no real non-trivial solutions, we have confirmed no real singular points for this model. 

Just as hoped, our results on Fig. 2  as well as the analytical confirmation show that this first model from the basis $\{R,R1\}$ has a late-time accelerating phase seen as a dynamical system attractor and is free from the separatrix problem and thus allows for a transition from a matter dominated phase to an accelerating late-time phase. 

\section{Model 3: $f(R,R1)=-\frac{m^{10}}{(-\frac{1}{3}R^2-8R1)^2}$}

Our next example in the basis $\{R,R1\}$ is the next higher power of the invariant of last section, namely the function
\be
f(R,R1)=-\frac{m^{10}}{(-\frac{1}{3}R^2-8R1)^2} 
\label{eq:GB1^2ModelA1=1/2}
\ee
for which we derive the generalized Friedmann equation 
\bea
\lefteqn{3H^2-\frac{m^{10}}{24(3\dot{H}^2+8H^2\dot{H}+8H^4)^4}(15\dot{H}^4+220H^2\dot{H}^3+672H^4\dot{H}^2+544H^6\dot{H}}\nonumber\\
& &+64H^8+60H\dot{H}^2\ddot{H}+160H^3\dot{H}\ddot{H}+96H^5\ddot{H})=0.
\label{eq:FriedmannGB1^2ModelA1=1/2}
\eea
\begin{figure}
\begin{center}
\begin{tabular}{|c|c|}
\hline
{\includegraphics[width=3.0in,height=3.0in,angle=0]{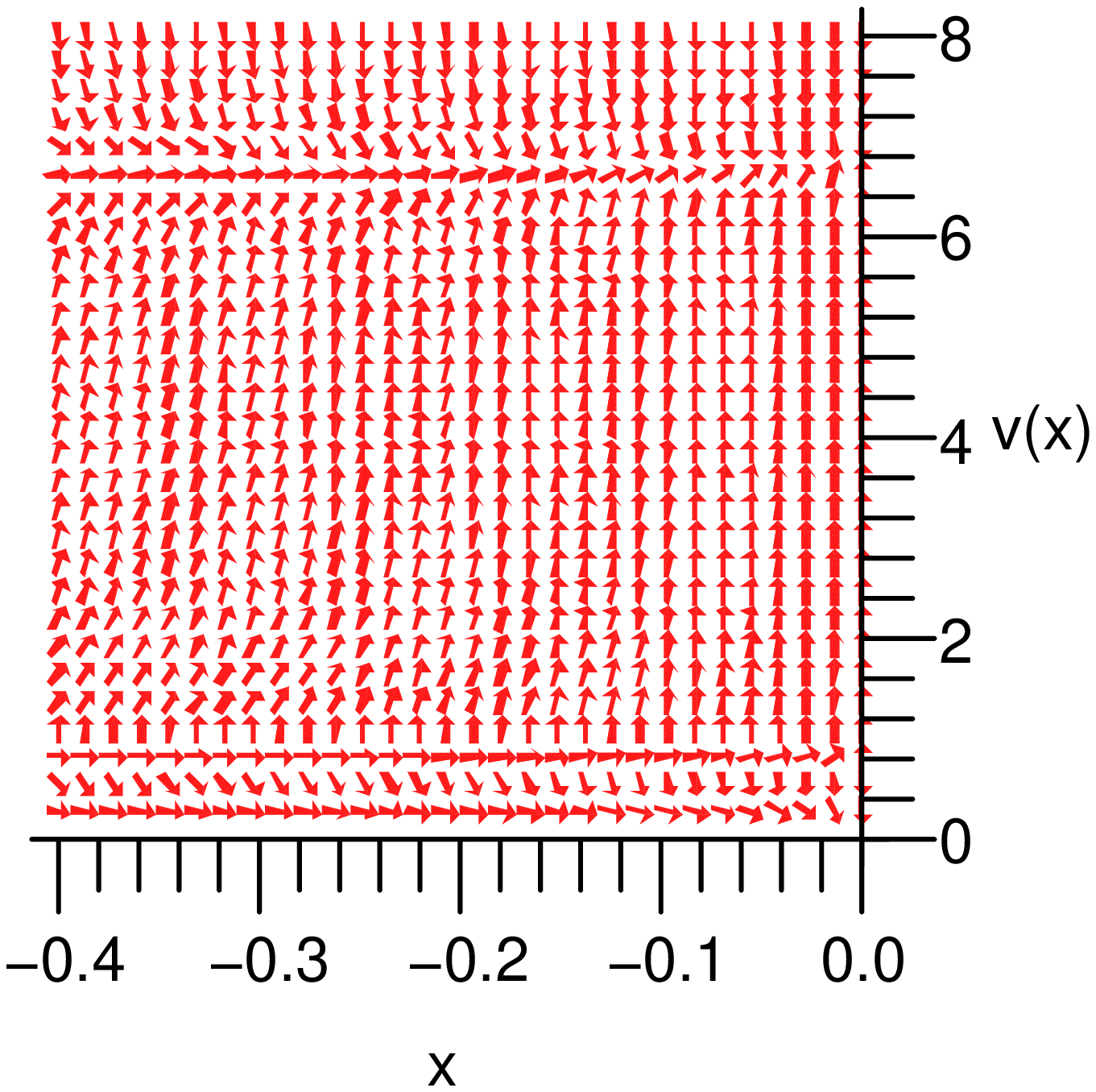}} &
{\includegraphics[width=3.0in,height=3.0in,angle=0]{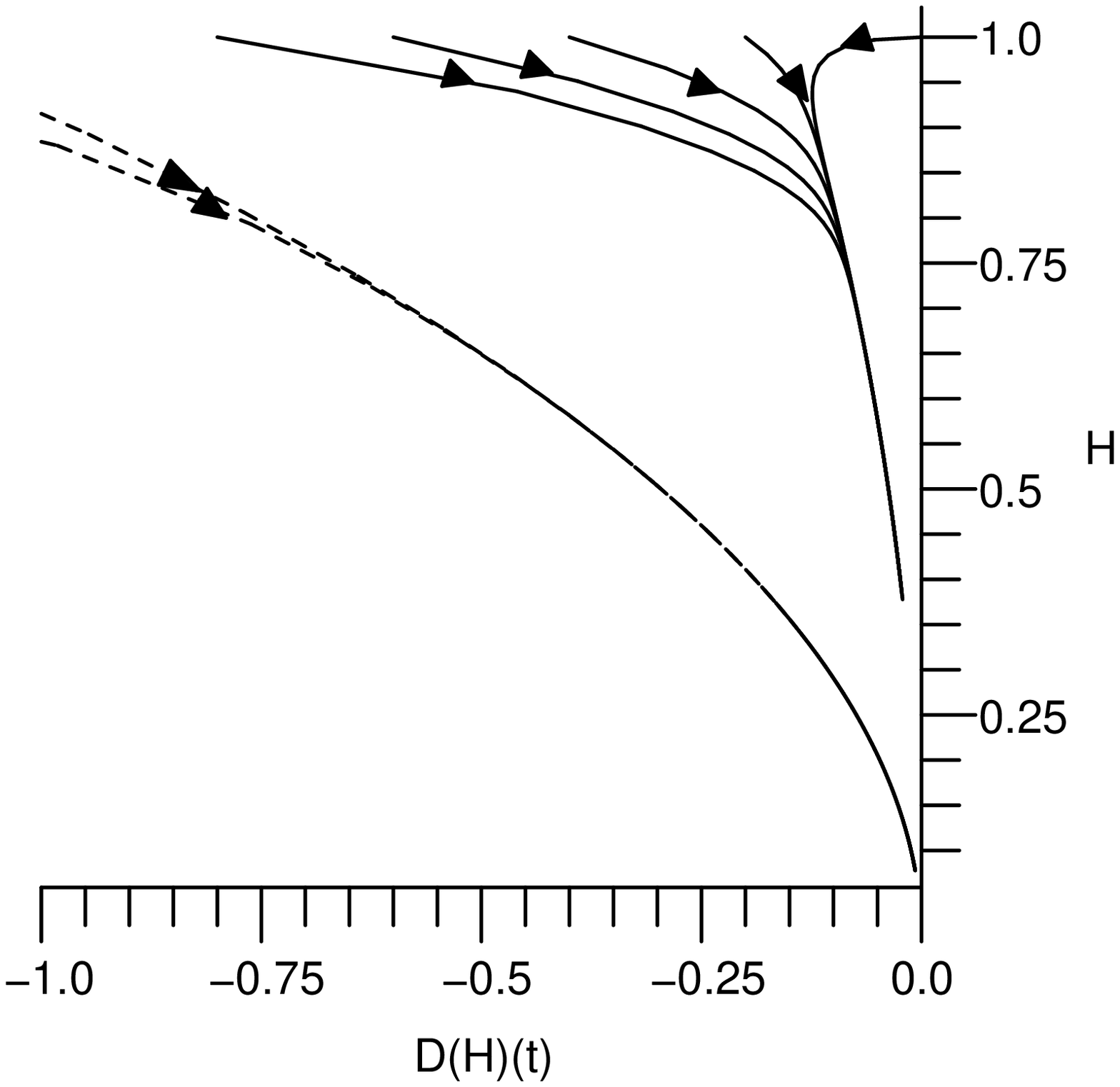}} \\
\hline
\end{tabular}
\caption{ 
LEFT: Dynamical systems for the case study $f(R,R1)=-\frac{m^{10}}{(-R^2/3-8R1)^2}$. In the first order phase portrait of the coordinate plane $(x,v)$, we find an accelerating attractor with the scale factor a(t) evolving as  $t^p$ where $p\approx6.6548 $.  There is also a non-accelerating solution of $p\approx0.8452$.
RIGHT: In the second order phase portrait with coordinates $(\dot{H},H)$, the solid-line branches are the accelerating ones with attractor having the scale factor a(t) as indicated on the left. Compared to Fig. 2, the solid-lines here are closer to the H(t) axis nearing the de Sitter solution (0,1). The dashed lines are non-accelerating. The horizontal axis on the right is $dH/dt$. 
} 
\end{center}
\end{figure}

The phase portrait for equation (\ref{eq:FriedmannGB1^2ModelA1=1/2}) in coordinates $(\dot{H},H)$ is plotted in Fig. 3. Once again, the condition for acceleration from above can be used to see which coordinate solutions (curves) will lead to the accelerating attractors (solid lines) and the non-accelerating solutions (dashed line).  The solutions look similar to those in Fig. 2, but the scalings have been changed in that the solutions approach de Sitter more rapidly.  The attractors (curves) have been pushed to the (right) H(t) axis by the more strongly driven accelerating dynamics of this model. This can be also seen in the following analytical examination. We analyze the Friedmann equation for these power-law solutions of $a(t)\propto{t^p}$ by parameterizing it in y(x) as before. Again, the earlier definitions (\ref{eq:v(x)parameter}) and (\ref{eq:diffv(x)parameter}) are employed to express (\ref{eq:FriedmannGB1^2ModelA1=1/2}) as 
\bea
x\frac{dv}{dx}&=&\frac{135v^4-540v^5+864v^6-544v^7+64v^8}{4v^3(15-40v+24v^2)}\nonumber\\
& &+\frac{(62208v-311040v^2+940032v^3-1870848v^4+2506752v^5-2211840v^6+1179648v^7-294912v^8-5832)x^{10}}{4m^{10}v^3(15-40v+24v^2)}\nonumber\\
& &
\label{eq:Friedmannv(x)GB1^2}
\eea

This first-order Friedmann equation is plotted in Fig. 3. In the phase space of $(x,v)$ we can see our attractors for this system. 
For our analytical confirmation that these are indeed attractors for late-time acceleration, we study the first-order Friedmann equation (\ref{eq:Friedmannv(x)GB1^2}) as $x\rightarrow{0}$,
\be
0=\frac{135-540v_{0}+864v_{0}^2-544v_{0}^3+64v_{0}^4}{4(15-40v_{0}+24v_{0}^2)}
\label{eq:GB1^2v0}
\ee
where we again eliminate the trivial solution. This allows us to find asymptotic future attractors in v(x) phase space which correspond to two real solutions $v_{0}=p=constant$, such that $v_{0}=15/4+3\sqrt{15}/4$ and $v_{0}=15/4-3\sqrt{15}/4$ as well as two complex solutions.  The solutions of interest for this study are power-law attractors corresponding to $p\approx 6.6548$ and $p\approx 0.8452$,  respectively.  Again, with at least one accelerating attractor with $p$ greater than 1, so we would need to assume some initial conditions to avoid the non-accelerating solution.  The attractor at $p\approx 6.6548$ is larger than the $2.9078$ obtained for the previous model confirming thus our interpretation above of the dynamical system plots indicating a stronger accelerating dynamics at late times for this model. We can extrapolate that even higher inverse powers of these first two models will have the larger of the two real solutions even more quickly approach the de Sitter solution. Also, upon performing a check to make sure that we do not have any singular points for this model we immediately notice that the nontrivial part of the denominator needed for study from the Friedmann equation has not changed from the previous model.  The solutions that cause the denominator of the Friedmann equation to blow up are imaginary.  Once again, we will have no real singular points for this model.  Therefore, this model has the desired self-accelerating late-time phase and is free from the separatrix problem, allowing thus for a passage from matter domination to cosmic acceleration.

\section{Model 4: $f(R,R1)=-\frac{m^{4n+2}}{\alpha[(a_1-\frac{5}{6})R^2-8R1]^n}$}

We generalize the two previous models using the following function of the minimal set $\{R,R1\}$ and write 
\be
f(R,R1)=-\frac{m^{4n+2}}{\alpha[(a_1-\frac{5}{6})R^2-8R1]^n}
\label{eq:GeneralActiona}
\ee
where $\alpha$ is a dimensionless constant, $n>0$ is an integer, and for which the general Friedmann equation reads
\bea
&&3H^2-\frac{m^{4n+2}}{2\alpha 6^n(6\beta\dot{H}^2+24\beta H^2\dot{H}+24\beta H^4-\dot{H}^2)^{2+n}}\Big{(}1152\beta^2H^6\dot{H}+576\beta^2nH^8-48\beta H^2\dot{H}^3-48\beta H^4\dot{H}^2\nonumber\\
&&-6nH^2\dot{H}^3+864\beta^2H^4\dot{H}^2+\dot{H}^4+288\beta^2nH^5\ddot{H}+576\beta^2n^2H^5\ddot{H}+4n^2H\dot{H}^2\ddot{H}+2nH\dot{H}^2\ddot{H}+792\beta^2nH^2\dot{H}^3\nonumber\\
&&+2448\beta^2nH^4\dot{H}^2+2592\beta^2nH^6\dot{H}-96\beta n^2H^2\dot{H}^3-192\beta n^2H^4\dot{H}^2-96\beta nH^2\dot{H}^3-48\beta nH^3\dot{H}\ddot{H}+576\beta^2H^8\nonumber\\
&&+72\beta^2nH\dot{H}^2\ddot{H}+144\beta^2n^2H\dot{H}^2\ddot{H}-24\beta nH\dot{H}^2\ddot{H}+288\beta^2nH^3\dot{H}\ddot{H}-24\beta n\dot{H}^4+72\beta^2n\dot{H}^4+36\beta^2\dot{H}^4-12\beta\dot{H}^4\nonumber\\
&&+2n\dot{H}^4-120\beta nH^4\dot{H}^2+2304\beta^2n^2H^6\dot{H}+48\beta nH^5\ddot{H}+288\beta^2H^2\dot{H}^3-48\beta n^2H\dot{H}^2\ddot{H}-96\beta n^2H^3\dot{H}\ddot{H}\nonumber\\
&&+576\beta^2n^2H^3\dot{H}\ddot{H}+576\beta^2n^2H^2\dot{H}^3+144\beta nH^6\dot{H}+2304\beta^2n^2H^4\dot{H}^2\Big{)}=0
\label{eq:GeneralFriedmanna}
\eea
where we have used $\beta=a1-5/6$ to simplify the notation. Recalling the earlier definitions to perform the dynamical systems analysis, we study if actions of this general type will lead to self-acceleration at late times. The generalized Friedmann equation becomes
\bea
&&x\frac{dv}{dx}=\Big{[}(-36\beta^2-8n^2-1+192\beta n^2v^2-48\beta v+288\beta^2v-864\beta^2v^2+48\beta v^2+72\beta n-216\beta^2n+1152\beta^2v^3-6nv\nonumber \\
&&-576\beta^2v^4+96\beta n^2-288\beta^2n^2+12\beta-6n-192\beta nv-3024\beta^2nv^2+1368\beta^2nv+2592\beta^2nv^3+144\beta nv^3\nonumber \\
&&+2304\beta^2n^2v^3-288\beta n^2v-3456\beta^2n^2v^2-576\beta^2nv^4+24\beta nv^2+1728\beta^2n^2v)+\Big{(}6-288\beta v^2+288\beta v\nonumber\\
&&+216\beta^2-72\beta-1728\beta^2v+5184\beta^2v^2-6912\beta^2v^3+3456\beta^2v^4\Big{)}\Big{(}\frac{\alpha\, x^2}{m^{4n+2}}\Big{)}\Big{(}\frac{6x^4(6\beta-24\beta v+24\beta v^2-1)}{v^2}\Big{)}^{n}\Big{]}\times\nonumber\\
&&\frac{-v}{2n(1+48\beta nv+24\beta v-144\beta^2v+144\beta^2v^2+288\beta^2nv^2+24\beta v^2-288\beta^2nv-24\beta n+36\beta^2+72\beta^2n-12\beta+2n)} \nonumber \\
\label{eq:GeneralFriedmannV(x)a}
\eea
which allows us to study the asymptotic behavior of the solutions to this equation at late times as seen in the specific examples $n=1$ and $n=2$ phase portraits of Figs. 2 and 3 respectively. We are looking for late time power-law solutions with the form $a(t)\propto t^p$, so we need to analytically solve for the general asymptotic solutions, $v(x)$ as $x\rightarrow{0}$ in equation (\ref{eq:GeneralFriedmannV(x)a}), i.e. 
\bea
&&0=(-36\beta^2-8n^2-1+192\beta n^2v_{0}^2-48\beta v_{0}+288\beta^2v_{0}-864\beta^2v_{0}^2+48\beta v_{0}^2+72\beta n-216\beta^2n+1152\beta^2v_{0}^3-6nv_{0}\nonumber\\
& &-576\beta^2v_{0}^4+96\beta n^2-288\beta^2n^2+12\beta-6n-192\beta nv_{0}-3024\beta^2nv_{0}^2+1368\beta^2nv_{0}+2592\beta^2nv_{0}^3+144\beta nv_{0}^3\nonumber\\
& &+2304\beta^2n^2v_{0}^3-288\beta n^2v_{0}-3456\beta^2n^2v_{0}^2-576\beta^2nv_{0}^4+24\beta nv_{0}^2+1728\beta^2n^2v_{0}) \times\nonumber\\
& &\frac{-v_{0}}{2n(1+48\beta nv_{0}+24\beta v_{0}-144\beta^2v_{0}+144\beta^2v_{0}^2+288\beta^2nv_{0}^2+24\beta v_{0}^2-288\beta^2nv_{0}-24\beta n+36\beta^2+72\beta^2n-12\beta+2n)}\nonumber\\
& &
\label{eq:V(x)GeneralSolutionA}
\eea
As usual, $H\rightarrow{0}$, $v_0\rightarrow{p}=$constant in equations (\ref{eq:v(x)parameter}) allows us to see that the asymptotic solutions of equation (\ref{eq:V(x)GeneralSolutionA}) have the general solutions 
\be
v_0=0, \frac{1}{2}\pm\frac{\sqrt{6}}{12\sqrt{\beta}}, \frac{12\beta+3n+42\beta n+48\beta n^2\pm\sqrt{\Xi}}{24\beta(1+n)},
\label{eq:V_0GeneralSolution1}
\ee
\be
where \,\,\Xi=240\beta n+900\beta^2n^2+9n^2+588\beta n^2+480\beta n^3+2880\beta^2n^3+2304\beta^2n^4+24\beta . 
\ee
As we will discuss below, these solutions contain at least one attractor with $p>1$ that represent a cosmological model with a late-time self accelerating phase. Recalling that we set $\beta=a1-5/6$, we can see that $a1=5/6$ is a singular point to be avoided. Now, if $a_1=1$ we return to the pure GB form with solutions $v_{0}=0,1,4n+1$ which suffer from a lack of transition from matter domination phase to an acceleration phase caused by separatrix. However, we are interested in more general models and need to further constrain these general solutions to avoid singular points that could be troublesome for cosmological transitions. 

Indeed, for a model to be cosmologically viable, separatrix points (singularities) must not exist between the epoch for nucleosynthesis (i.e. $p=1/2$) and the transition from matter domination to an accelerating phase (i.e. $p=1$). For that, we analyze the denominator of the generalized Friedmann equation (\ref{eq:GeneralFriedmanna}) searching for singular points.  For power law expansion, i.e. $a(t)=t^p$, the denominator of (\ref{eq:GeneralFriedmanna}) vanishes for 
\be
\frac{p^2(6\beta-24\beta p+24\beta p^2-1)}{t^4}=0
\label{eq:GeneralFriedmannaZeroP}
\ee
with the non-trivial solutions given by 
\be
p_{1,2}=\frac{1}{2}\pm\frac{\sqrt{6}}{12\sqrt{\beta}}.
\label{eq:GeneralZero}
\ee

These are also the first pair of solutions to the general dynamical system approach seen earlier in (\ref{eq:V_0GeneralSolution1}). If the solutions (\ref{eq:GeneralZero}) are real then they constitute singular points for the models. If their values are $1/2 \le p \le 1$ then they represent the undesired separatrix. 

Now, $\beta$ is negative for $a_1<5/6$ and equation (\ref{eq:GeneralZero}) has no real values thus no singular points. This will then avoid the separatrix problem for the generalized models of equation (\ref{eq:GeneralActiona}). In this case, the last two solutions of equations (\ref{eq:V_0GeneralSolution1}) represent the possible attractors. For $n=1$, in the range $5/6>a_1>367/507+10\sqrt{3}/169$ these are negative attractors of no-interest. In the range $367/507+10\sqrt{3}/169\ge a_1\ge 367/507-10\sqrt{3}/169$ there are no real last solutions.  And finally, in range $367/507-10\sqrt{3}/169>a_1$ will give two real positive attractors with one large enough for self-acceleration as shown in the two previous examples.  In the case, $n=2$, the boundaries of the three ranges change to $5/6>a_1>153/196+1\sqrt{85}/196$, $153/196+1\sqrt{85}/196\ge a_1\ge 153/196-1\sqrt{85}/196$, and $153/196-1\sqrt{85}/196>a_1$, as separated in the $n=1$ case, respectively.  Finally, for the general $n$ case, the new boundary terms can be found from
\be
a_1=\frac{1}{3n^2(25+80n+64n^2)}\Big(38n^2+180n^3+160n^4-10n-1\pm \sqrt{\Sigma}\Big),
\label{eq:A1SeparatrixA}
\ee
\be
with \,\,\Sigma=1+20n+149n^2+530n^3+944n^4 +800n^5+256n^6,
\label{eq:A1SeparatrixB}
\ee
where $(+)$ and $(-)$ solutions will give ranges in the same way as cases $n=1,2$.
Next, for $a_1>5/6$, the solutions $p_{1,2}$ are real singular points and their exact values must be outside the interval $[1/2,1]$ in order to avoid separatrix problems. In this case, the previous dynamical solutions (\ref{eq:V_0GeneralSolution1}) are all real allowing for at least one self-accelerating model with $p>1$.

\section{Model 5: $f(R,R1)=(\frac{1}{3}R^2+8R1)\exp{\frac{1}{-\frac{1}{3}R^2-8R1}}$}

Using some guidance from previous studies of $f(R)$ models \cite{Amendola2006}, we continue the study of inverse functions of the invariant $(-R^2/3-8R1)$ in the basis $\{R,R1\}$ by considering the addition of an exponential function, 
\be
f(R,R1)=(\frac{1}{3}R^2+8R1)\exp{\frac{1}{-\frac{1}{3}R^2-8R1}},
\label{eq:ActionexpGB1}
\ee
to the action.  The variation of our exponential action with respect to the metric, again yields, after some algebra, the Friedmann equation as
\bea
\lefteqn{3H^2-\frac{1}{3(3\dot{H}^2+8H^2\dot{H}^3+8H^4)^3}[e^{\Sigma}(1458H^2\dot{H}^7-80H^4\dot{H}^2+396H^4\dot{H}^4+48H^6\dot{H}^3-212544H^8\dot{H}^4+576H^{10}\dot{H}}\nonumber\\
& &-10368H^4\dot{H}^6-1008H^5\dot{H}^2\ddot{H}-1152H^7\dot{H}\ddot{H}-82944H^{11}\dot{H}\ddot{H}-42768H^5\dot{H}^4\ddot{H}-89856H^7\dot{H}^3\ddot{H}-432H^3\dot{H}^3\ddot{H}\nonumber\\
& &-11664H^3\dot{H}^5\ddot{H}-1458H\dot{H}^6\ddot{H}-114048H^9\dot{H}^2\ddot{H}-9H\dot{H}^2\ddot{H}-81H\dot{H}^4\ddot{H}-24H^3\dot{H}\ddot{H}-83376H^6\dot{H}^5-64H^6\dot{H}\nonumber\\
& &+297H^2\dot{H}^5-27648H^{13}\ddot{H}-16H^5\ddot{H}-24H^2\dot{H}^3-300672H^{10}\dot{H}^3-235008H^{12}\dot{H}^2-82944H^{14}\dot{H}\nonumber\\
& &-576H^9\ddot{H}+81\dot{H}^6+729\dot{H}^8+768H^{12})]=0
\label{eq:FriedmannexpGB1}
\eea
for convenience we have used $\Sigma=-\frac{1}{6(3\dot{H}^2+8H^2\dot{H}^3+8H^4)}$. The phase portrait in coordinates $(\dot{H},H)$ is not shown because it does not have any real solutions as we shall see by applying the phase-space parameters, in the same way we have used in previous models. The Friedmann equation takes the form
\bea
\lefteqn{x\frac{dv}{dx}=-v[e^{\tilde{\Sigma}}((-440640v^4+466560v^5+21870v+82944v^7-99144v^2-2187-290304v^6+263088v^3)x^8}\nonumber\\
& &+(2256v^5-1620v^4-576v^7-1152v^6-81v^2+567v^3+768v^8)x^4+72v^5+64v^7-112v^6-18v^4)+(-7128v^4\nonumber\\
& &+13824v^7-19008v^6-4608v^8-243v^2+14976v^5+1944v^3)x^6]\Big{/}[e^{\tilde{\Sigma}}((114048v^4+27648v^6-11664v+42768v^2\nonumber\\
& &-82944v^5-89856v^3+1458)x^8+(-432v^3+81v^2-1152v^5+576v^6+1008v^4)x^4+16v^6+9v^4-24v^5)]
\label{eq:Friedmannv(x)expGB1}
\eea
where $\tilde{\Sigma}=-\frac{x^4(3-8v+8v^2)}{6v^2}$. The phase portrait in Fig. 4 for the phase space $(x,v)$ shows that there are no late-time power-law attractors. Analytically, we confirm this result with the asymptotic solution to equation (\ref{eq:Friedmannv(x)expGB1}) as $x\rightarrow{0}$
\be
0=\frac{72v_{0}+64v_{0}^3-112v_{0}^2-18}{16v_{0}^2+9-24v_{0}}
\label{eq:v_0expGB1}
\ee

\begin{figure}
\begin{center}
\begin{tabular}{|c|c|}
\hline
{\includegraphics[width=3.0in,height=3.0in,angle=0]{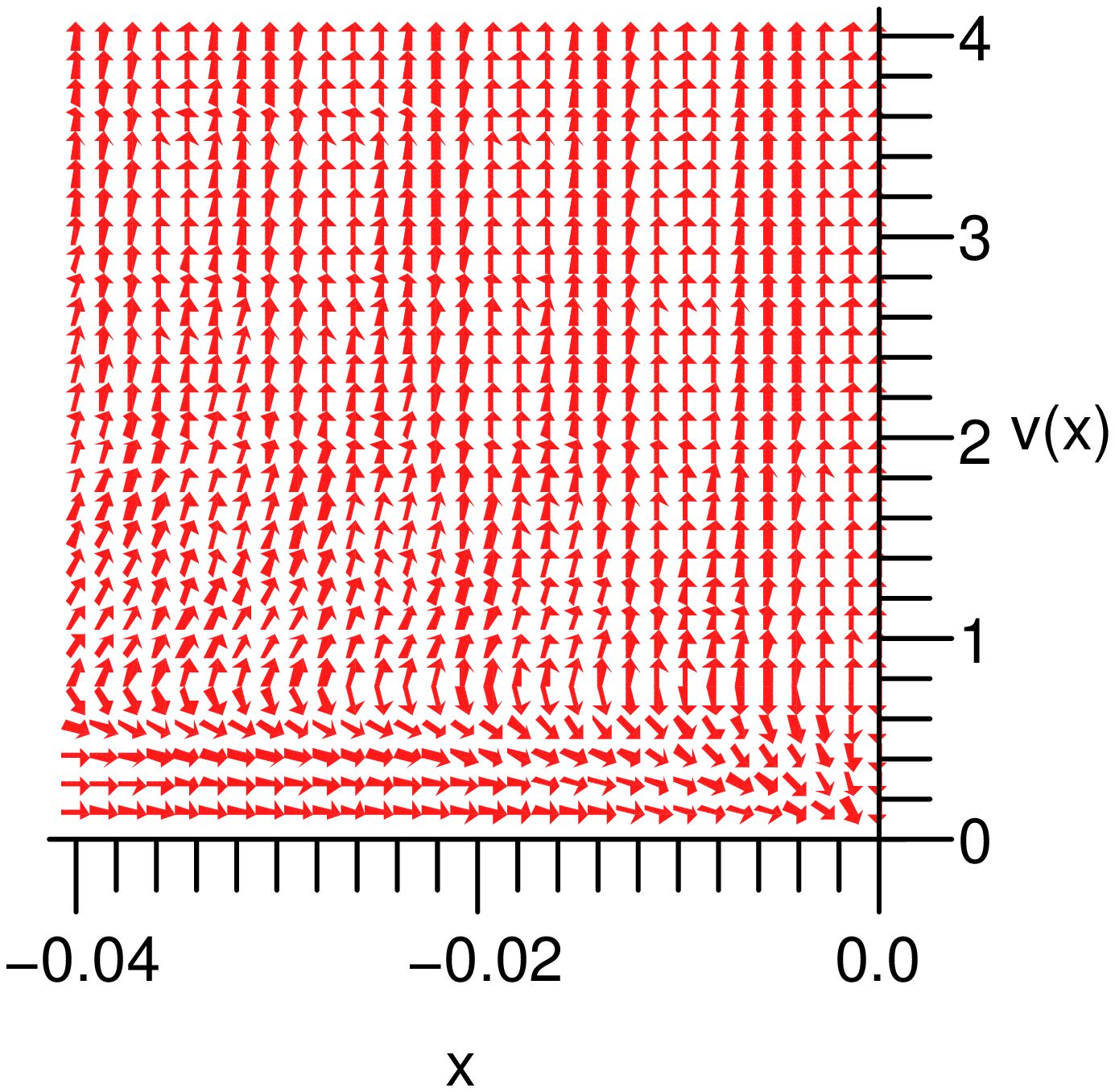}}  
& {\includegraphics[width=3.0in,height=3.0in,angle=0]{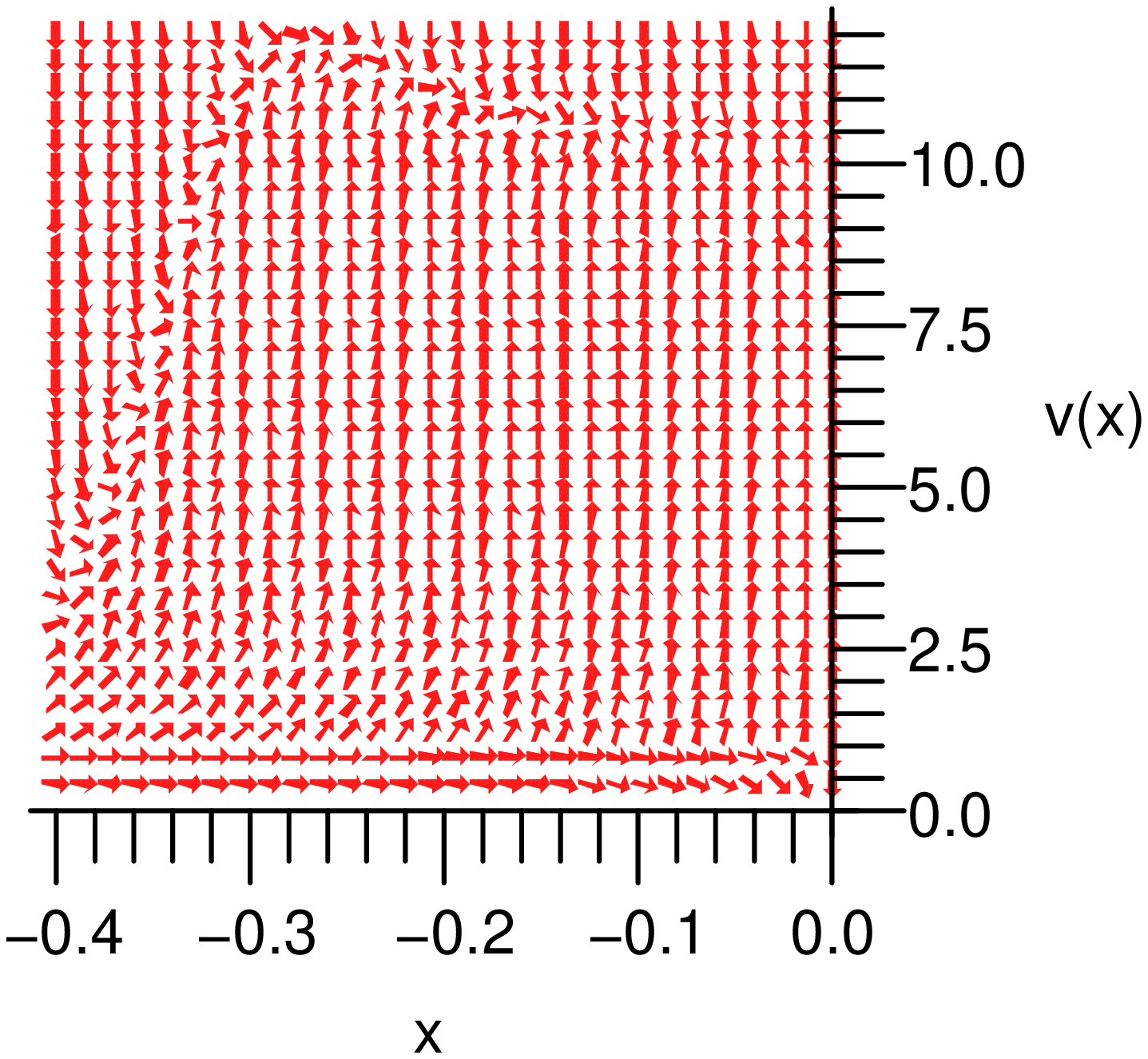}}\\
\hline
\end{tabular}
\caption{\label{fig:} 
LEFT: Dynamical systems for the case study $f(R,R1)=(R^2/3+8R1)\exp{\frac{1}{-R^2/3-8R1}}$. In the first order phase portrait of the coordinate plane $(x,v)$, we see no evidence of an accelerating attractor with the scale factor a(t) evolving as  $t^p$ where $v_{0}=p=constant$. This behavior is explained in section IX. RIGHT: Dynamical systems for the case study of the expansion of the exponential action  carried out to the first three terms $f(R,R1)=-(1+\frac{1}{-R^2/3-8R1}+\frac{1}{2(-R^2/3-8R1)^2}+\frac{1}{6(-R^2/3-8R1)^3})$.  Here the accelerating attractors from the $n=3$ case dominate the series action at $p=10.500$ and $p\approx0.8125$ however as explained in section IX when $n$ goes to infinity the solution becomes a pure de Sitter one.
} 
\end{center}
\end{figure}

where we eliminate the trivial solution. Which if we had power-law attractors, they would come from real solutions to equation (\ref{eq:v_0expGB1}), but no real solutions exist, and thus, the exponential action does not lead to late-time power-law accelerating dynamics. It leads to a pure de Sitter inflationary phase. This lack of real solutions can best be understood by studying the behavior of the pure exponential action in a series expansion added to the original Einstein-Hilbert action.  We begin with the usual series
\be
\exp{z}=1+z+\frac{1}{2}z^2+\frac{1}{6}z^3+\frac{1}{24}z^4+O(z^5)
\label{eq:EXPSeries}
\ee
where we use $z=1/(-R^2/3-8R1)$ and keep only up to two terms in $z$ for simplification to derive 
\be
f(R,R1)=-\left(1+\frac{1}{-R^2/3-8R1}+\frac{1}{2(-R^2/3-8R1)^2}\right)
\label{eq:SeriesAction}
\ee
We see from this series action, that at cosmological scales (long distances), the higher order term will dominate.  In this action, the dominate term is represented by Model 4 from the previous section where $n=2$, and in fact if we solve this action in the usual way, to find the accelerating attractors, we arrive at the two complex and two real solutions from Model 4, where the real solutions at $v_0=15/4\pm3\sqrt{15}/4$ are the same as those plotted in Fig. 2.  We can extrapolate here, as we have done previously, that adding higher terms to the action from the original series expansion, such as the third term from equation (\ref{eq:EXPSeries}), will increase the value of $n$, so that the action becomes
\be
f(R,R1)=-\left(1+\frac{1}{-R^2/3-8R1}+\frac{1}{2(-R^2/3-8R1)^2}+\frac{1}{6(-R^2/3-8R1)^3}\right)
\label{eq:SeriesAction2}
\ee
and the dominant term is represented by the $n=3$ case, so that equation (\ref{eq:V_0GeneralSolution1}) yields the real solutions of $v_0=181/32\pm155/32$ shown in Fig. 3.  As we add more and more terms, we return to the pure exponential model above with an infinite number of terms, meaning $n\rightarrow{\infty}$. Now, looking at the solutions given in (\ref{eq:V_0GeneralSolution1}) and taking their limit when $n\rightarrow{\infty}$, we see that in the last two solutions, the larger solution tends to infinity, or pure de Sitter (as for example noted for $f(R)$ models in \cite{Amendola2006}) and the smaller solution goes to zero.  This only leaves the other pair of solutions there, depending only on $\beta$, not $n$, which due to $a_1<5/6$ give complex solutions just as we saw for the exponential action and the asymptotic equation (\ref{eq:v_0expGB1}). Multiplying the infinite series by $(-R^2/3-8R1)$ or $1/(-R^2/3-8R1)$ will only slow this behavior by one term, or expedite it, respectively, neither of which change the overall result for the exponential action. 

In sum, the analysis of the series confirms the result from Fig. 4 and equations (\ref{eq:v_0expGB1}) that the action with the exponential term go to a pure de Sitter behavior. 

The same techniques can be applied to study the behavior of a logarithmic action of the form $(-R^2/3-8R1)\ln{(-R^2/3-8R1)}$ where the expansion can be written as 
\be
ln(z)=\frac{x-1}{x}+\frac{1}{2}\frac{x-1}{x}^2+\frac{1}{3}\frac{x-1}{x}^3+O(z^4)
\label{eq:LNSeries1}
\ee
which by collecting a certain number of terms can be simplified to the form
\be
ln(z)=\frac{11}{6}-\frac{3}{x}+\frac{3}{2x^2}-\frac{1}{3x^3}
\label{eq:LNSeries2}
\ee
where we can see a logarithmic action of the form from \cite{Amendola2006} can be shown to have similar behavior to the exponential action on cosmological scales as we add more higher terms to return the series to its pure infinite form also with a pure de Sitter phase.
\section{Conclusion}

We discussed a systematic approach to higher-order gravity models based on a connection made between higher-order models and theorems on minimal sets of independent invariants as derived from invariants theory in general relativity. Various theorems demonstrate how for a given type of symmetry and source of the spacetime, there exist a unique basis of curvature invariant in terms of which the other invariants can be written. 

We found that the approach allows one to narrow the basic building elements to two invariants $\{R,R1\}$ in the case of interest covering all the Friedmann-Lemaitre-Robertson-Walker manifolds. 

We showed that as a result of the independent nature of the invariants used, the approach allows one to avoid any undesired redundancies in the dynamical equations.  

We investigated some examples built from the basis $\{R,R1\}$ using dynamical system analyses supplemented by analytical calculations and found that some combinations have cosmological evolution with late-time self-accelerating phase and do allow for a passage from a matter dominated phase to a cosmic acceleration phase. 

Another class of models that need to be considered using the basis approach is that of models with strong coupling between curvature invariants and scalar fields, see for example \cite{Nojiri2005,Lobo2008,Neupane2007,KoivistoMota2007} and references therein. 

Finally, the models discussed here need to be subjected to further physical acceptability constraints and comparisons with astronomical data at solar system level and cosmological scales. Such a study will be presented elsewhere \cite{MoldenhauerIshak2008}.  

\acknowledgments 
The authors thank Wolfgang Rindler and Damian Easson for useful comments. MI acknowledges partial support from the Hoblitzelle Foundation and a Clark award.


\begin{thebibliography}{}
%
\bibitem{observations} 

A. G. Riess, {\em{et al.}}, Astron. J. {\textbf{116}}, 1009-1038 (1998);
%
 S. Perlmutter, {\em{et al.}}, Astrophys. J. {\textbf{517}}, 565-586 (1999);
%
R. A. Knop, {\em{et al.}}, Astrophys. J. {\textbf{598}}, 102-137
(2003); 
%
A. G. Riess, {\em{et al.}}, Astrophys. J. {\textbf{607}}, 665-687
(2004);  
%
C. L. Bennett, {\em{et al.}},
Astrophys. J. Suppl. Ser. {\textbf{148}}, 1 (2003); 
D. N. Spergel, {\em{et al.}},
Astrophys. J. Suppl. Ser. , 175 (2003);
L. Page et al., 
Astrophys. J. Suppl. Ser. {\textbf{148}}, 2333 (2003).
%
 Seljak et al., Phys.Rev. D{\textbf{71}}, 103515 (2005);
%
M. Tegmark, {\em{et al.}},  Astrophys. J. {\textbf{606}}, 702-740
(2004).
%
D.N. Spergel , {\em{et al.}}, 
Astrophys.J.Suppl. 170, 377 (2007).

\bibitem{reviews} S. Weinberg
Rev. Mod. Phys., 
\textbf{61}, 1 (1989);
%
M.S. Turner, 
\physrep, 
\textbf{333}, 619 (2000);
%
V. Sahni, A. Starobinsky
Int.J.Mod.Phys. 
\textbf{D9}, 373 (2000);
%
S.M. Carroll, 
Living Reviews in Relativity, 
\textbf{4}, 1 (2001);
%
T. Padmanabhan, 
\physrep, 
\textbf{380}, 235 (2003);
%
P.J.E. Peebles and B. Ratra, 
Rev. Mod. Phys. 
\textbf{75}, 559 (2003);
%
A. Upadhye, M. Ishak, P. J. Steinhardt, 
Phys. Rev. D
\textbf{72}, 063501 (2005);
%
A. Albrecht et al, {\it Report of the Dark Energy Task Force}
astro-ph/0609591 (2006).
%
M. Ishak, 
Foundations of Physics Journal, 
Vol.\textbf{37}, No 10, 1470 (2007).  
%
\bibitem{Biblio300}We conducted a thourough bibliography search on NASA ADS system, HEP-SPIRES and arXiv:[astro-ph][gr-qc] archives. 
%
\bibitem{Carroll2005} S. M. Carroll, A. De Felice, V. Duvvuri, D. A. Easson, M. Trodden, M. S. Turner, Phys. Rev. D \textbf{71} 063513 (2005).
%
\bibitem{Easson2004} D.A. Easson, Int. J. Mod. Phys. A19, 5343 (2004).
%
\bibitem{Easson2005} D. A. Easson, F. P. Schuller, M. Trodden, M. N.R. Wohlfarth, Phys.Rev. D72 043504 (2005). 
%
\bibitem{Lobo2008} F. S. N. Lobo, arXiv:0807.1640v1 [gr-qc] (2008).
%
\bibitem{NojiriOdintsov2006a} S. Nojiri, S. D. Odintsov, Int.J.Geom.Meth.Mod.Phys. 4 (2007) 115-146.
%
\bibitem{Sotiriou2006} T. P. Sotiriou, Class. Quant. Grav. \textbf{23},  1253 (2006).
%
\bibitem{MengWang2004} X. Meng, P. Wang, Class. Quant. Grav. \textbf{21}, 2029 (2004).
%
\bibitem{NojiriOdintsov2003} S. Nojiri, S. D. Odintsov, Phys.Rev. D68 (2003) 123512.
%
\bibitem{Mena2006} O. Mena, J. Santiago, J. Weller, Phys. Rev. Lett. \textbf{96}, 041103 (2004).
%
\bibitem{Shirata2005} A. Shirata, T. Shiromizu, N. Yoshida, Y. Suto, Phys. Rev. D \textbf{71} 064030 (2005).
%
\bibitem{Borowiec2006} A. Borowiec, W. Godlowski, M. Szydloski, Phys. Rev. D \textbf{74} 043502 (2006).
%
\bibitem{NavarroVanAcoleyen2005} I. Navarro, K. Van Acoleyen, Phys. Lett. B \textbf{622} 1 (2005).
%
\bibitem{Chiba2003} T. Chiba, Phys. Lett. B \textbf{575} 1 (2003).
%
\bibitem{Chiba2007} T. Chiba, T. L. Smith, A. L. Erickcek, Phys. Rev. D \textbf{75} 124014 (2007).
%
\bibitem{Faraoni2006a} V. Faraoni, Phys. Rev. D \textbf{74} 023529 (2006).
%
\bibitem{NojiriOdintsov2007} S. Nojiri, S. D. Odintsov, 	arXiv:0710.1738v2 (2007).
%
\bibitem{DologovKawasaki2003} A.D. Dolgov, M. Kawasaki, Phys. Lett. B \textbf{573} 1 (2003).
%
\bibitem{Faraoni2006b} V. Faraoni, Phys. Rev. D \textbf{74} 104017 (2006).
%
\bibitem{Sotiriou2007} T. P. Sotiriou, Ph. D. Thesis, arxiv:0710.4438 (2007).
%
\bibitem{BirrellDavies1982} N.D. Birrell and P.C.W. Davies, \textit{Quantum Fields in Curved Space} (Cambridge University Press 1982).
%
\bibitem{Stelle1977} K. S. Stelle, Phys. Rev. D \textbf{16} 953 (1977).
%
\bibitem{Debever1964} R. Debever, Cah. Phys. \textbf{168} 303 (1964).
%
\bibitem{CarminatiMcLenaghan1991}  J. Carminati, R. McLenaghan, J. Math. Phys. \textbf{32}(11) 3135 (1991).
%
\bibitem{ZakharyMcIntosh1997} E. Zakhary C. McIntosh, Gen. Rel. Grav. \textbf{29} (5) 539 (1997).
%
\bibitem{Segre}
C. Segre, {\textit{Memorie della R. Accademia dei Lincei, ser. 3a XIX, 127, Sec 5.1}} (1884). 
%
\bibitem{Petrov1954} A. Z. Petrov, Uch. Zapiski Kazan Gos. Univ. \textbf{144} (1954);  Reprint: Gen. Rel. Grav. \textbf{32} 16665 (2000).
%
\bibitem{Pirani1957} F. E. A. Pirani, Phys. Rev. \textbf{105} 1089 (1957).
%
\bibitem{Penrose1960} R. Penrose, Ann. Phys. \textbf{10} 171 (1960). 
%
\bibitem{Stephani2003} H. Stephani, D. Kramer, M. MacCallum, C. Hoenselaers, E. Herlt \textit{Exact Solutions of Einstein's Field Equations} (2nd edn.) (Cambridge University Press 2003).
%
\bibitem{Thomas1934} T. Y. Thomas, \textit{The Differential Invariants of Generalized Spaces} (Cambridge University Press 1934).
%
\bibitem{NarlikarKarmarkar1948} V. V. Narlikar, K. R. Karmarkar, Proc. Ind. Acad. Sci. A \textbf{29} 91 (1948).
%
\bibitem{GeheniauDebever1956} J. Geheniau, R. Debever, Bull. Cl. Sci. Acad. R. Belg. \textbf{XLII} 114 (1956).
%
\bibitem{Debever1956} R. Debever, Bull. Cl. Sci. Acad. R. Belg. \textbf{XLII} 252 (1956).
%
\bibitem{Witten1959} L. Witten, Phys. Rev. \textbf{113} 357 (1959).
%
\bibitem{Petrov1969} A. Z. Petrov, \textit{Einstein Spaces} (Pergamon, Oxford 1969).
%
\bibitem{Santosuosso1999} G. L. Santosuosso, \textbf{35} (7) 1307 (1999).  
%
\bibitem{CarminatiZakhary1999} J. Carminati, E. Zakhary, Class. Quant. Grav. \textbf{16} 3221 (1999).
%
\bibitem{Chiba2005} T. Chiba, JCAP \textbf{0503} 008 (2005).
%
\bibitem{Dvali2006} G. Dvali, New J.Phys. \textbf{8} 326 (2006).
%
\bibitem{LiBarrowMota2007} B. Li, J. D. Barrow, D. F. Mota, Phys. Rev. D \textbf{76} 044027 (2007).
%
\bibitem{NavarroVanAcoleyen2006} I. Navarro, K. Van Acoleyen, JCAP 0603 008 (2006).
%
\bibitem{DeFelice2006} A. De Felice, M. Hindmarsh, M. Trodden, JCAP \textbf{0608} 005 (2006).
%
\bibitem{Rindler2006} W. Rindler, \textit{Relativity:Special, General, and Cosmological, Second Edition} (Oxford University Press 2006).
%
\bibitem{Amendola2006} L. Amendola, R. Gannouji, D. Polarski, S. Tsujikawa, Phys. Rev. D \textbf{75} (2007) 083504.
%
\bibitem{Nojiri2005} S. Nojiri, S. D. Odintsov, M. Sasaki, Phys. Rev. D \textbf{71} (2005) 123509.
%
\bibitem{Neupane2007} B. M. N. Carter and I. P. Neupane, JCAP \textbf{0606}, (2006) 004.  
%
\bibitem{KoivistoMota2007} T. Koivisto, D. F. Mota
 Phys.Lett.B \textbf{644}, (2007) 104-108;
%
\bibitem{MoldenhauerIshak2008}J. Moldenhauer, M. Ishak, in preparation (2008).
%


\end{thebibliography}
\end{document}